\documentclass{article}
\usepackage{latexsym,amssymb}
\newtheorem{prop}{Proposition}
\newtheorem{lemma}{Lemma}

\newcommand{\R}{\mathbb{R}}
\newcommand{\N}{\mathbb{N}}

\def\open#1{\setbox0=\hbox{$#1$}
\baselineskip = 0pt
\vbox{\hbox{\hspace*{0.4 \wd0}\tiny $\circ$}\hbox{$#1$}} 
\baselineskip = 10pt\!}

\begin{document} 

\noindent {\sc Comm.\ Math.\ Sci.}\hfill
\copyright\ 2004 International press\\
\noindent Vol.\ 2, No. 2, pp.\ 145--158

\bigskip

\begin{center}
{\Large \bf
GLOBAL WEAK SOLUTIONS TO THE RELATIVISTIC VLASOV-MAXWELL SYSTEM REVISITED
\footnote{Received: March 2, 2004; accepted (in revised version): 
April 22, 2004. Communicated by Francois Golse}
\\}
\ \\
{\large GERHARD REIN
\footnote{University of Bayreuth, Department of Mathematics,
        D-95440 Bayreuth, Germany
        (gerhard.rein@uni-bayreuth.de)} }\\

\end{center}

{\bf Abstract.}
In their seminal work \cite{DL1}, R.~DiPerna and P.-L.~Lions established
the existence of global weak solutions to the Vlasov-Maxwell system.
In the present notes we give a somewhat simplified proof of this
result for the relativistic version of this system, the main purpose being to
make this important result of kinetic theory more easily
accessible to newcomers in the field. We show that the weak solutions
preserve the total charge.

\bigskip


\section{Introduction}
\setcounter{equation}{0}
When a plasma is sufficiently rarefied and/or sufficiently hot
like in the solar wind or in a powered-up fusion reactor collisions
among the plasma particles are sufficiently rare to be neglected.
The only interaction among the particles then is through the
electromagnetic fields which the particles create collectively.
For the sake of simplicity we restrict ourselves to a plasma
consisting of just one particle species, say, electrons,
and we allow for the possibility that the particles move at relativistic
speeds. The time evolution of the plasma
is governed by the relativistic Vlasov-Maxwell system:
\begin{equation} \label{vlasov}
\partial_t f + \widehat{p} \cdot \partial_x f + 
\left( E + \widehat{p} \times B\right)\cdot \partial_p f = 0,
\end{equation}
\begin{equation} \label{maxev}
\partial_t E - \mathrm{curl}\, B = - 4 \pi j,\ \ 
\partial_t B + \mathrm{curl}\, E = 0,
\end{equation}
\begin{equation} \label{constr}
\mathrm{div}\, E = 4 \pi \rho,\ \ \mathrm{div}\, B = 0,
\end{equation}  
\begin{equation} \label{rhojdef}
\rho(t,x) = \int f(t,x,p)\, dp,\ \ 
j(t,x) = \int \widehat{p} f(t,x,p)\, dp.
\end{equation}
Here $f=f(t,x,p)$ denotes the density of the particles
on phase space, $t\in\R,\ x,p \in \R^3$ stand for time, 
position, and momentum,
\[
\widehat{p} := \frac{p}{\sqrt{1+|p|^2}}
\]
is the velocity of a particle with momentum $p$,
$E=E(t,x)$ and $B=B(t,x)$ are the electromagnetic fields,
$\rho = \rho(t,x)$ and $j=j(t,x)$ denote the spatial charge 
density and current,
and units are chosen such that all physical constants such as the speed
of light and the charge and rest mass of an individual particle are
normalized to unity. The analysis can immediately
be adapted to a plasma with several species of particles. 

We are interested in the Cauchy problem for the above
system, i.e., in the existence of solutions satisfying the initial
conditions
\begin{equation} \label{incond}
f_{|t=0} = \open{f}\,,\ E_{|t=0} = \open{E}\,,\ B_{|t=0} = \open{B}\,
\end{equation}
where the initial data satisfy the constraint part (\ref{constr}) 
of the Maxwell
equations. Global existence and uniqueness of sufficiently smooth
solutions
to this initial value problem is an open problem. 
Local existence and uniqueness of classical solutions
for smooth, compactly supported data was established in \cite{GlSt1}. 
These solutions can be extended
globally in time provided the momentum support can be controlled,
which has been done for data which are small \cite{GlSt2}
or close to neutral \cite{GlSch1}
or close to spherically symmetric \cite{R1}.
In lower dimensions global classical solutions exist for general data
\cite{GlSch2,GlSch3,GlSch4,GlSch5}. Different approaches to the 
result in \cite{GlSt1} were recently given in \cite{BG,KS}.
There is as yet no indication that classical solutions for general data 
in three dimensions develop singularities. Nevertheless, it is natural
to weaken the solution concept in order to obtain global solutions.
This was done by R.~DiPerna and {P.-L.}~Lions \cite{DL1}.
The authors
restricted themselves to the non-relativistic Vlasov-Maxwell system 
where $\widehat{p}$
is replaced by $p$ and pointed out that their arguments apply 
to the relativistic case as well.
Many results on the Vlasov-Maxwell
system are reviewed in the monograph \cite{Gl1}, in particular, global
weak solutions are discussed following \cite{DL1}, but using also
arguments introduced in \cite{Kr}.
The techniques in \cite{DL1} are closely related
to those used by the same authors in their seminal work on the 
Boltzmann equation \cite{DL2}, and later these techniques have been
adapted to a variety of other problems in kinetic theory;
we mention \cite{CR,KRST} as being related to the present notes.

Given the fact that the global existence and uniqueness problem
for the Vlasov-Maxwell system in three dimensions is still open
and the fact that in recent years many young researchers have joined
the field it should be useful to give a simplified proof of the
result of DiPerna and Lions.
We emphasize
that all the {\em essential} techniques we are going to use 
are introduced in \cite{DL1}, but some non-trivial concepts 
and arguments from \cite{DL1} are avoided.
We concentrate on the relativistic version of the system, firstly,
because being Lorentz invariant it seems better justified from 
a physics point of view, and secondly, because
the non-relativistic case was considered in detail in \cite{DL1,Gl1}.
In passing we note that there are global existence results
for classical solutions of the Vlasov-Poisson system \cite{Pf,LP,Sch}
but not yet for the so-called relativistic Vlasov-Poisson
system where $p$ is replaced by $\widehat{p}$ in the Vlasov equation, 
cf.\ \cite{GlSch}.

We now discuss how the paper and the proof proceed and how our
version differs from the original one. In the next section we recall
various a-priori bounds resulting from conservation of energy and
conservation of phase space volume by the characteristic flow
of the Vlasov equation. Then we introduce a regularized version
of the system which has global in time, smooth
solutions. It will be important that these regularized
solutions exist on the whole time axis. In \cite{DL1} the system
was regularized by adding a sufficiently large power of the Laplacian
to the evolution part of the Maxwell equations (\ref{maxev}). 
This destroys the
time reversibility of the system and technically impedes the straight
forward application of the velocity averaging lemma discussed below.
We propose to regularize the system by smoothing the current $j$
so that conceptually we remain closer to the unmodified system. 
This regularization is due to
\cite{Ho} and was used in the context of weak solutions
in \cite{Gl1,Kr,KR}, but its technical advantages
were not fully realized. Given a sequence of solutions to regularized 
Vlasov-Maxwell systems along which the regularization vanishes in the limit
we show that the a-priori bounds hold uniformly. 
Hence we can extract a weakly convergent subsequence whose limit
is the candidate for the desired weak solution of the unmodified system.
The main difficulty lies in passing to the limit in the nonlinear
term in the Vlasov equation. This difficulty can be overcome since 
additional compactness of the approximating sequence is provided by the
velocity averaging lemma. Its application is discussed in Section~5,
and to make the present notes self-contained we give a proof of the
relevant relativistic version in an appendix. Since the proof rests
on Fourier transforming the Vlasov equation with respect to space and time
it is advantageous that our approximating solutions are defined on the whole
time axis. In the set-up of \cite{DL1} a certain cut-off and extension
maneuver was necessary. 
A second problem lies in passing to the limit in 
moments of $f$ like $\rho$ and $j$, the difficulty being that
the relevant weights in $p$ are not test functions. 
As opposed to \cite{DL1}
we derive weak convergence of $\rho$ and $j$ directly
from a-priori bounds for these quantities. 
In all this a minimal requirement is that
the initial data have finite energy and finite total charge.
The original proof assumed in addition that $\open{f}\,$ is square integrable.
We assume that $\open{f}\,$ is bounded, and we completely avoid
the non-trivial concept of renormalization which was necessary in 
\cite{DL1}---in the opinion of the present author the main motivation 
for studying weak solutions is {\em not} to allow for the greatest 
possible generality
in the initial data but the lack of global existence results
for stronger solution concepts.
Having obtained a weak solution we examine some of its properties 
in Section~6. Sufficient regularity is established
to make sense of saying that the solution satisfies the initial conditions.
Then we prove that for almost
all times $t$ the total charge and more generally any $L^q$-norm
of $f(t)$ equals its initial value, a result
which was not obtained in \cite{DL1} and for which we exploit the
relativistic nature of the system. 
Since the Vlasov equation is a conservation law
on phase space it is a desirable feature of any ``reasonable''
solution concept that solutions preserve the total charge.
The energy at times $t\neq 0$ is bounded
by its initial value but is not known to be conserved,
and neither are weak solutions known to be unique.

To sum up, our proof is simplified compared to \cite{DL1}
in the sense that the only non-trivial tool employed is velocity
averaging. The rest of the proof consists of straight forward exploitation
of straight forward a-priori bounds. 

\section{Preliminaries and a-priori bounds} 
\setcounter{equation}{0}
We introduce the main conservation laws
for the relativistic Vlasov-Maxwell system; the derivations are formal,
and in which sense they hold depends on the type of solution
under consideration. Writing the Vlasov equation (\ref{vlasov})
in divergence form
\begin{equation} \label{vlasovdiv}
\partial_t f + \mathrm{div}_x\left[\widehat{p} \, f\right] + 
\mathrm{div}_p\left[\left( E(t,x) + 
\widehat{p} \times B(t,x)\right)\,f \right] = 0
\end{equation}
and integrating with respect to $p$ yields local conservation of charge,
\begin{equation} \label{lchargecons}
\partial_t \rho + \mathrm{div}_x j = 0.
\end{equation}
On the level of the solutions of the characteristic system
\[
\dot x = \widehat{p},\ \ \dot p = E(t,x) + \widehat{p} \times B(t,x)
\]
of the Vlasov equation, conservation of charge is reflected in 
the induced flow on phase space being measure preserving
and $f$ being constant along the flow.
A lengthy computation shows local conservation of energy,
\begin{equation} \label{lenergycons}
\partial_t e + \mathrm{div}_x \sigma = 0,
\end{equation}
where the corresponding energy density and flux are defined by
\begin{eqnarray*}
e(t,x) 
&:=&
\int \sqrt{1+|p|^2} f(t,x,p)\,dp + \frac{1}{8\pi} 
\left( |E(t,x)|^2 + |B(t,x)|^2 \right),\\
\sigma (t,x)
&:=&
\int p\, f(t,x,p)\, dp + \frac{1}{4\pi} E(t,x) \times B(t,x).
\end{eqnarray*}
These local conservation laws imply
corresponding global conservation laws:
\begin{prop} \label{cq}
Consider a classical solution of the relativistic Vlasov-Maxwell
system with compactly supported
initial data $\open{f}\, \in C^1_c(\R^6)$, $\open{f}\, \geq 0$, 
$\open{E}\,, \open{B}\, \in C^2_c(\R^3)$,
satisfying the constraints (\ref{constr}).
As long as the solution exists its energy
\[
\int\!\!\!\!\!\int \sqrt{1+|p|^2} f(t,x,p)\,dp\,dx + 
\frac{1}{8 \pi} \int \left(|E(t,x)|^2 + |B(t,x)|^2\right)\, dx
\]
as well as any  $L^q$-norm $\|f(t)\|_q$ with $1\leq q \leq \infty$
are constant in time.
\end{prop}
We use the conserved quantities to derive bounds on $\rho$
and $j$: For any $R>0$, 
\begin{eqnarray*}
\rho(t,x)
&=&
\int_{|p|\leq R}f(t,x,p)\,dp + \int_{|p|> R}
f(t,x,p)\,dp\\
&\leq& 
\frac{4\pi}{3} R^3 \|f(t)\|_{\infty}  +
R^{-1}\int \sqrt{1+|p|^2} f\,dp
\leq C \left(\int \sqrt{1+|p|^2}f\,dp\right)^{3/4},
\end{eqnarray*}
where for the last step we choose
\[
R=\left(\int \sqrt{1+|p|^2}f\,dp\right)^{1/4},
\]
and the constant depends on $\|\open{f}\,\|_\infty$.
Taking both sides of the estimate to the power $4/3$, integrating in $x$,
and using Proposition~\ref{cq} we have the following a-priori bounds
on $\rho$ and hence also $j$ which is dominated by $\rho$:
\begin{prop} \label{rhojbounds}
Along any solution as considered in Proposition~\ref{cq},
\[
\|\rho(t)\|_{4/3},\; \|j(t)\|_{4/3} \leq C
\]
where the constant $C$ depends on the energy of the initial data and on
$\|\open{f}\,\|_\infty$.
\end{prop}
To conclude we observe that the constraints (\ref{constr}) propagate:
If we have a solution of the system 
(\ref{vlasov}), (\ref{maxev}), (\ref{rhojdef}) 
satisfying the constraints (\ref{constr}) initially then
the constraints hold as long as the solution exists, since
\[
\partial_t(\mathrm{div}_x E - 4 \pi \rho) =
\mathrm{div}_x \partial_t E - 4 \pi \partial_t \rho = 
\mathrm{div}_x (\partial_t E + 4 \pi j)
= \mathrm{div}_x (\mathrm{curl}_x B) = 0,
\]
\[
\partial_t \mathrm{div}_x B = - \mathrm{div}_x (\mathrm{curl}_x E) = 0.
\]  

\section{The regularized system} \label{sregsys}
\setcounter{equation}{0}
For a mollifier
\begin{equation} \label{smoother}
d \in C^\infty_c (\R^3),\ d \geq 0,\ \int d = 1,\ d\ \mbox{even}
\end{equation}
we consider the regularized relativistic Vlasov-Maxwell system 
(\ref{vlasov}), (\ref{maxev}), (\ref{rhojdef}) where in the 
Maxwell equations
(\ref{maxev}) we replace $j$ by $d \ast j$, the convolution referring to $x$;
recall that the constraints (\ref{constr}) propagate once they hold initially.
Along a (local) solution of the regularized system charge is still conserved, 
hence $\int |j(t)| \leq \int \rho(t) \leq C$ and therefore all spatial 
derivatives of $d\ast j$ are bounded uniformly in $t$. 
This is sufficient to 
show that the regularized system has global-in-time solutions for 
initial data as specified
in Proposition~\ref{cq}; details of the proof can be found in \cite{Ho}. 
However, there is one difficulty with the regularized system: 
The time derivative of the energy is
\[
\int\left(j \cdot E - (d\ast j) \cdot E\right)\,dx
\]
which need not vanish.
Hence, when constructing a sequence of approximating solutions by 
the regularization above we have to make sure  
we preserve the a-priori bounds, uniformly along the sequence.
Let 
\[
L^1_\mathrm{kin} (\R^6) := \left\{ g\in L^1(\R^6) \mid 
g\geq 0,\ \int\!\!\!\!\!\int \sqrt{1+|p|^2} g(x,p)\, dp\, dx < \infty \right\}
\]
be endowed with the weighted $L^1$-norm $\|\cdot\|_\mathrm{kin}$
with weight $\sqrt{1+|p|^2}$.
We fix initial data
\begin{equation} \label{data}
\open{f}\, \in L^1_\mathrm{kin}\cap L^\infty (\R^6),\ 
\open{E}\,,\ \open{B}\, \in L^2(\R^3)
\end{equation}
satisfying the constraint (\ref{constr}) in the sense of distributions, and
take sequences $(\open{f}\,_n) \subset C^\infty_c(\R^6)$, 
bounded in $L^\infty(\R^6)$, and 
$(\open{E}\,_n^\sim),(\open{B}\,_n^\sim) \subset C^\infty_c(\R^3)$ such 
that for any $p\in[1,\infty[$,
\[
\open{f}\,_n \to \open{f}\, \ \mbox{in} 
\ L^1_\mathrm{kin}\cap L^p (\R^6),\quad
\open{E}\,_n^\sim \to \open{E}\,,\ \open{B}\,_n^\sim \to \open{B}\, 
\ \mbox{in} \ L^2(\R^3).
\]
With $d$ as above we let $d_n (x) := n^{3} d(n x)$, define
\[
\open{E}\,_n := d_n \ast \open{E}\,_n^\sim \to \open{E}\,,\
\open{B}\,_n := d_n \ast \open{B}\,_n^\sim \to \open{B}\, 
\ \mbox{in} \ L^2(\R^3),
\]
and denote by $(f_n,E_n,B_n)$ the global solution of the regularized
system with the initial data $(\open{f}\,_n,\open{E}\,_n,\open{B}\,_n)$,
where $j$ now is replaced by $d_n \ast d_n\ast j_n$---the two $d_n$'s
are intentional---with $j_n$ defined
in terms of $f_n$ as in (\ref{rhojdef}). This solution exists
for all $t \in \R$ by the reasoning above. By uniqueness,
\[
(E_n,B_n)(t) = d_n \ast (E_n^\sim,B_n^\sim)(t) 
\]
where $(E_n^\sim,B_n^\sim)$ solves the Maxwell equations (\ref{maxev})
with initial data $\open{E}\,_n^\sim,\open{B}\,_n^\sim$ and
current $d_n \ast j_n$. The energy
\[
\int\!\!\!\!\!\int \sqrt{1+|p|^2} f_n(t,x,p)\,dp\,dx + 
\frac{1}{8 \pi} \int \left(|E_n^\sim (t,x)|^2 + |B_n^\sim(t,x)|^2\right)\, dx
\] 
now {\em is} constant in time, its time derivative becoming
\[
\int\left(j_n \cdot E_n - (d_n\ast j_n) \cdot E_n^\sim\right)\,dx =0.
\]
Since the modified energy defined above dominates
the energy of the regularized solution we can use the arguments in
the previous section to prove the following result, 
cf.\ Proposition~\ref{rhojbounds}:
\begin{prop} \label{approxbounds}
For initial data as specified in (\ref{data}) there exists a constant $C>0$
such that for any $n\in \N$ the solution $(f_n,E_n,B_n)$
of the regularized initial value
problem described above exists for all $t\in \R$ and satisfies the a-priori
bounds
\[
\|f_n(t)\|_\mathrm{kin},\ \|f_n(t)\|_\infty,\
\|E_n(t)\|_2,\ \|B_n(t)\|_2,\ \|\rho_n(t)\|_{4/3},\ \|j_n(t)\|_{4/3} \leq C.
\]
\end{prop}

\section{The weak limit} 
\setcounter{equation}{0}
\begin{prop} \label{weaklimit}
There exist functions 
\[
f\in L^\infty(\R;L^1_\mathrm{kin}\cap L^\infty (\R^6)),\ 
E, B \in L^\infty (\R; L^2(\R^3)),\
\rho, j \in  L^\infty (\R; L^{4/3}(\R^3))
\]
such that up to a subsequence
\[
f_n \rightharpoonup f\ \mbox{in}\ L^2(I \times\R^{6}),\
E_n,B_n \rightharpoonup E,B\ \mbox{in}\ L^2(I \times\R^{3}),\
\rho_n,j_n \rightharpoonup \rho,j\ \mbox{in}\ L^{4/3}(I \times\R^{3})
\]
for any bounded interval $I \subset \R$,
\[
f\geq 0,\ \rho = \int f\,dp,\ j = \int \widehat{p} f\,dp
\ \mbox{a.~e.,}
\]
and the Maxwell equations (\ref{maxev}), (\ref{constr}) as well as
local conservation of charge (\ref{lchargecons}) hold in the sense
of distributions. For almost all $t\in \R$ the energy of $(f(t),E(t),B(t))$
is bounded by its initial value.
\end{prop}
\noindent {\bf Proof.\ }
The extraction of the weakly convergent subsequence is standard,
and by a diagonal sequence argument this subsequence can be chosen
independently of the interval $I$. Since $f_n \geq 0$ for $n\in \N$,
the weak limit $f$ is non-negative almost everywhere.
That the limits lie in the asserted
function spaces is straight forward.
For example let $A\subset \R$ be bounded and measurable with
Lebesgue measure $\lambda(A)$, and let $R>0$. Then
\begin{eqnarray*}
&&
\int_A\int_{|x|\leq R}\left(\int_{|p|\leq R}\sqrt{1+|p|^2} f\,dp 
+ \frac{1}{8\pi}(|E|^2 + |B|^2) \right)
\,dx\,dt\\
&&\quad\leq\liminf_{n\to \infty}
\int_A\int_{|x|\leq R}\left(\int_{|p|\leq R}\sqrt{1+|p|^2} f_n\,dp 
+ \frac{1}{8\pi}(|E_n|^2 + |B_n|^2) \right)
\,dx\,dt\\
&&\quad
\leq \lambda(A) \left(\|\open{f}\,\|_\mathrm{kin} 
+ \frac{1}{8\pi}\left(\|\open{E}\,\|_2^2 + \|\open{B}\,\|_2^2\right) \right).
\end{eqnarray*}
Since $R>0$ and $A\subset \R$ are arbitrary the assertion 
on the energy follows, which implies part of the assertion 
on the weak limits $f,E,B$.

For the moments $\rho$ and $j$ we can argue similarly, 
but we have to make sure that the
weak limit of $\rho_n$ is the density $\rho$ induced by $f$ and analogously
for $j$. To this end, consider a test function 
$\psi \in C^\infty_c(\R\times \R^3)$
and some $R>0$. Then
\begin{eqnarray*}
&&
\int\!\!\!\!\!\int \left(\int \widehat{p} f dp - j\right)\, \psi \,dx\, dt\\
&& 
\qquad =
\int\!\!\!\!\!\int \left(\int_{|p|\leq R} \widehat{p} f dp - j\right)
\, \psi \,dx\, dt
+
\int\!\!\!\!\!\int\!\!\!\!\!\int_{|p|> R} \widehat{p} f\, dp\, \psi \,dx\, dt\\
&&
\qquad =
\lim_{n\to \infty}
\int\!\!\!\!\!\int \left(\int_{|p|\leq R} \widehat{p} f_n dp - j_n\right)
\, \psi \,dx\, dt
+ 
\int\!\!\!\!\!\int\!\!\!\!\!\int_{|p|> R} 
\widehat{p} f\, dp\, \psi \,dx\, dt \\
&&
\qquad =
\lim_{n\to \infty}\int\!\!\!\!\!\int\!\!\!\!\!\int_{|p|> R}
\widehat{p}\,(f-f_n)\, dp \,\psi \,dx\, dt.
\end{eqnarray*}
The modulus of the latter integral can be estimated by
\[
\|\psi\|_\infty \frac{1}{R} 
\int\!\!\!\!\!\int_{\mathrm{supp}\, \psi}
\int \sqrt{1+|p|^2} (f_n + f)\,dp\,dx\, dt \leq \frac{C}{R}
\]
via the uniform bound on the kinetic energy,
and since $R>0$ and the test function $\psi$ 
are arbitrary the assertion for $j$
follows. That local conservation of charge (\ref{lchargecons}) and the 
second of the Maxwell equations (\ref{maxev})
hold in the sense of distributions is obvious. As to the first of the Maxwell
evolution equations let $\psi$ be as above. Then with the abbreviation 
$\delta_n := d_n\ast d_n$,
\[
\int\!\!\!\!\!\int \delta_n \ast j_n \psi\,dx\,dt 
- \int\!\!\!\!\!\int j \psi\,dx\,dt 
= 
\int\!\!\!\!\!\int (\delta_n \ast j_n - j_n)\psi\,dx\,dt 
+ \int\!\!\!\!\!\int (j_n - j)\psi\,dx\,dt.
\] 
The second term converges to zero, 
and the first can be estimated by
\[
\|j_n\|_{L^{4/3}(\mathrm{supp}\, \psi)} \|\delta_n\ast \psi - \psi\|_4
\]
which converges to zero as well. 
Hence $\delta_n\ast j_n \to j$ in the sense of distributions.
The computation in Section~2 which showed that
the constraints (\ref{constr}) propagate can be performed in the weak sense,
and the proof is complete. \hspace*{\fill} $\Box$

\section{Compactness via momentum averaging} 
\setcounter{equation}{0}

In the present section we want to show that the weak limit
obtained in the previous one satisfies the Vlasov equation
in the sense of distributions. Due to the non-linearity in the
latter equation this is the crucial problem in the whole proof.
We need to show that for any $\phi \in C^\infty_c(\R \times \R^6)$,
\[
\int\!\!\!\!\!\int\!\!\!\!\!\int (E_n + \widehat{p} \times B_n)
\cdot \partial_p \phi\, f_n
dp\,dx\,dt 
\to
\int\!\!\!\!\!\int\!\!\!\!\!\int (E + \widehat{p} \times B)
\cdot \partial_p \phi\, f\,dp\,dx\,dt,
\]
possibly after extracting a further subsequence. By a well known density 
argument it is sufficient to consider test functions which
factorize: $\phi(t,x,p) = \phi_1(t,x) \phi_2(p)$ with test
functions $\phi_1, \phi_2$. The desired convergence will follow once
we can show that for any bounded open interval $I \subset \R$,
any $S>0$, and any $\psi \in  C^\infty_c(\R^3)$ up to a subsequence,
\begin{equation} \label{momavconv}
\int \psi(p)\, f_n(\cdot,\cdot,p)\, dp \to 
\int \psi(p)\, f (\cdot,\cdot,p)\,dp \ \mbox{strongly in}\ L^2(I \times B_S),
\end{equation}
where $B_S \subset \R^3$ denotes the ball of radius $S$ about the origin.
That this is true is the consequence of the velocity-averaging lemma,
a version of which was established in \cite{DL1}, cf.~also \cite{GLPS}, 
and which for obvious reasons we prefer to call momentum-averaging lemma. 
We state it here and give a proof
in an appendix:
\begin{lemma} \label{momav}
Let $R>0$ and $\psi \in  C^\infty_c(B_R)$. There exists a constant 
$C>0$ such that for any functions 
$h,g_0,g_1 \in L^2(\R \times \R^3\times B_R)$ 
which satisfy the inhomogeneous transport equation
\begin{equation} \label{transport}
\partial_t h + \widehat{p} \cdot \partial_x h = g_0 + \mathrm{div}_p g_1
\end{equation}
in the sense of distributions we have
\[
\int \psi(p) h (\cdot,\cdot,p)\,dp \in H^{1/4}(\R \times \R^3)
\]
with 
\[
\left\|\int \psi(p) h (\cdot,\cdot,p)\,dp\right\|_{H^{1/4}}
\leq C 
\left(\|h\|_{L^2(\R \times \R^3\times B_R)}
+ \|g_0\|_{L^2(\cdots)} + \|g_1\|_{L^2(\cdots)}\right)
\]
\end{lemma}
Here $H^{1/4}$ denotes the usual fractional order Sobolev space
defined in terms of the Fourier transform. In the proof one needs to
Fourier transform the transport equation with respect to
$t$ and $x$ so that it is essential that the equation holds 
for all $t \in \R$. Since we apply the lemma
to the approximating solutions we avoid considerable technical
complications by having the latter defined on the whole real line.

In order to prove (\ref{momavconv}) we take an arbitrary
open, bounded interval $I \subset \R$, 
$\psi \in C^\infty_c (\R^3)$, and $R>0$ such that 
$\mathrm{supp}\, \psi \subset B_R$. Moreover, we choose some test function
$\zeta \in C^\infty_c(\R)$ such that $0\leq \zeta \leq 1$
and $\zeta_{|I} = 1$. Then for $\tilde f_n (t,x,p) := \zeta(t) f_n(t,x,p)$
the Vlasov equation in divergence form (\ref{vlasovdiv}) implies that
\[
\partial_t \tilde f_n + \widehat{p}\cdot\partial_x \tilde f_n 
= g_0 + \mathrm{div}\,_p g_1
\]
where $g_0 := \zeta' f_n$ and 
$g_1 := (E_n + \widehat{p} \times B_n) \zeta f_n$.
By Proposition~\ref{approxbounds} the $L^2$-norms of the latter
functions over the domain $\R \times \R^3\times B_R$ are bounded,
uniformly in $n$. Hence by Lemma~\ref{momav} we conclude that
the momentum averages of the functions $\tilde f_n$, formed with the
test function $\psi$, lie in the Sobolev space
$H^{1/4}(\R \times \R^3)$, and their corresponding $H^{1/4}$-norms
are bounded, uniformly in $n$. Over the open, bounded set
$I\times B_S$ with $S>0$ arbitrary this Sobolev space is compactly
embedded in $L^2$, by choice of $\zeta$ the functions $\tilde f_n$
coincide with $f_n$ on $I\times B_S$, and hence along a suitable subsequence
Eqn.~(\ref{momavconv}) is established.

We conclude that the weak limit $(f,E,B)$ obtained in the previous 
section satisfies the complete
relativistic Vlasov-Maxwell system on $\R \times \R^3 \times \R^3$
in the sense of distributions.

\section{Continuity properties and conservation laws} 
\setcounter{equation}{0}

In order to be justified in saying that the weak solution
$(f,E,B)$ satisfies the initial conditions we need to establish
some minimal continuity in $t$. We consider this problem for $f$,
the arguments for $E$ and $B$ being very similar. We express 
$\partial_t f_n$ via the Vlasov equation in divergence form, integrate in 
time, multiply the
result by a test function $\phi \in C^\infty_c(\R^6)$ and integrate
in $x$ and $p$ to obtain, after an integration by parts:
\begin{eqnarray*}
&&
\int\!\!\!\!\!\int f_n(t,x,p)\, \phi(x,p)\, dp\,dx
=
\int\!\!\!\!\!\int \open{f}\,_n(x,p)\, \phi(x,p)\, dp\,dx \\
&&
\qquad \qquad + \int_0^t\int\!\!\!\!\!\int
\left[\widehat{p} \cdot \partial_x \phi + 
(E_n + \widehat{p} \times B_n)\cdot \partial_p \phi \right]
f_n(s,x,p)\, dp \, dx\, ds.
\end{eqnarray*}
For $t\in \R$ we define a distribution 
$\tilde f(t)\in {\cal D}'(\R^6)$ by
\begin{eqnarray*}
\langle \tilde f (t), \phi \rangle
&:=&
\int\!\!\!\!\!\int \open{f}\, (x,p)\, \phi(x,p)\, dp\,dx\\
&&
{} + \int_0^t\int\!\!\!\!\!\int
\left[\widehat{p} \cdot \partial_x \phi + (E + \widehat{p} \times B)
\cdot \partial_p \phi\right]
f (s,x,p)\, dp \, dx\, ds.
\end{eqnarray*}
The mapping $t \mapsto \tilde f(t) \in {\cal D}'(\R^6)$
is continuous in the sense of distributions, $\tilde f(0) = \open{f}\,$,
and one can show that this is a representative of $f$.
By a density argument
$t\mapsto \tilde f(t)$ is also continuous with respect
to the weak topology of $L^q(\R^6)$ for any $1< q < \infty$. 
Since for any bounded measurable set $M\subset \R^6$ and $s>0$
the negative order Sobolev space $W^{-s,2}(M)$ is compactly embedded
in $L^2(M)$ continuity with respect to 
the negative order Sobolev norm follows.

We already noted that the total energy can at least not increase, and that
local conservation of charge (\ref{lchargecons})
holds in the sense of distributions. It is simple to conclude that
for almost all $t$, $\int \rho (t)$ equals some fixed constant, but
we assert that this constant
really is the initial charge. Indeed, more is true:
\begin{prop} \label{lpcons}
The weak solution obtained above preserves all $L^q$-norms of $f(t)$,
more precisely, for every $1\leq q \leq \infty$,
\[
\|f(t)\|_q = \|\open{f}\,\|_q \ \mbox{for a.~a.}\ t\in \R.
\]
Moreover, $t\mapsto f(t)$ is a.~e.\ strongly $L^q$ continuous in the
following sense: There exists a set of continuity $C \subset \R$ such that
the Lebesgue measure
$\lambda(\R\setminus C) = 0$, $0\in C$, and for any $q\in ]1,\infty[$
the mapping
$C \ni t \mapsto f(t) \in L^q (\R^6)$
is strongly continuous.
\end{prop}
\noindent {\bf Proof.\ }
Since $\partial_t \rho_n + \mathrm{div}\, j_n = 0$ and 
$|j_n| \leq \rho_n$ we have for every $R>0$ and $t>0$,
\begin{eqnarray*}
\frac{d}{dt} \int_{|x| > R + t} \rho_n(t)\, dx 
&=&
- \int_{|x| = R + t} \rho_n(t)\, dS_x + \int_{|x| > R + t} 
\partial_t\rho_n(t)\, dx\\
&=&
- \int_{|x| = R + t} \rho_n(t)\, dS_x - \int_{|x| > R + t} 
\mathrm{div}\, j_n (t)\, dx\\
&=&
- \int_{|x| = R + t} (\rho_n(t) + \nu \cdot j_n (t))\, dS_x \leq 0
\end{eqnarray*}
where $\nu$ is the outer unit normal of the domain $\{|x| > R + t\}$. 
The analogous argument works for $t<0$ and the domain $\{|x| > R - t\}$. 
Hence 
\begin{equation} \label{finprop}
\int_{|x| > R + |t|} \rho_n (t)\, dx \leq \int_{|x| > R} \open{\rho}_n\, dx,\ 
t\in \R,\ R>0,\ n\in \N,
\end{equation}
where $\open{\rho} := \int \open{f}\, \, dp$. We claim that
\begin{equation} \label{chargeconsae}
\int \rho (t)\, dx =  \int\open{\rho}\, dx\ \mbox{for a.~a.}\ t \in \R.
\end{equation}
Let $\epsilon >0$ be arbitrary.
Since $\open{\rho}$ is integrable, we can choose $R>0$ 
such that 
\[
\int_{|x| > R} \open{\rho}\, dx < \epsilon.
\]
By the convergence of the initial data
and (\ref{finprop}) we conclude that
\[
\int_{|x| > R + |t|} \rho_n\, dx \leq \int_{|x| > R} \open{\rho}_n\, dx 
< \epsilon
\]
for all $t\in \R$ and all sufficiently large $n\in \N$. 
Let $A \subset \R$ be measurable and bounded. Then
\begin{eqnarray*}
\int_A \int \rho\, dx\,dt
&\geq&
\int_A \int_{|x| \leq R + |t|} \rho\, dx\,dt
=
\lim_{n \to \infty} \int_A \int_{|x| \leq R + |t|} \rho_n\, dx\,dt\\
&=&
\lim_{n \to \infty} \int_A \left[\int \rho_n\, dx - 
\int_{|x| > R + |t|} \rho_n\, dx\right] dt >
\lambda(A)\left[\int \open{\rho}\, dx - \epsilon \right],
\end{eqnarray*}
and for sufficiently large $S>0$ we have by monotone convergence,
\begin{eqnarray*}
\int_A \int \rho\, dx\,dt
&\leq&
\int_A \int_{|x| \leq S} \rho\, dx\,dt + \lambda(A) \epsilon \\
&=&
\lim_{n \to \infty}
\int_A \int_{|x| \leq S} \rho_n\, dx\,dt + \lambda(A) \epsilon 
\leq
\lambda(A) \left( \int \open{\rho}\, dx + \epsilon \right).
\end{eqnarray*}
This implies
that there exists a set $M_\epsilon \subset \R$ of measure zero 
such that
\[
\int \open{\rho}\, dx -  \epsilon \leq \int \rho(t)\, dx 
\leq \int \open{\rho}\, dx + \epsilon,\ 
t \in \R \setminus M_\epsilon.
\]
Hence (\ref{chargeconsae}) holds on $\R \setminus \cup_{k\in \N}M_{1/k}$.
For general $q\in [1,\infty[$ we define
\[
\rho_q (t,x):= \int f^q(t,x,p)\, dp,\ 
j_q (t,x):= \int \widehat{p} f^q(t,x,p)\, dp,
\]
with analogous definitions for $\rho_{q,n}$ and $j_{q,n}$.
Since $f$ and $f_n$ are bounded, these new densities converge 
in the same sense as the ones for $q=1$ and the weak limits are 
indeed the $q$-densities induced by $f$.
Moreover, $\partial_t \rho_{q,n} + \mathrm{div}\, j_{q,n} = 0$ classically 
for every $n$, and $|j_{q,n}| \leq \rho_{q,n}$. Hence exactly the 
same argument as in the case
$q=1$ shows that
\begin{equation} \label{qchargeconsae}
\int \rho_q (t)\, dx =  \int\open{\rho}_q\, dx
\end{equation}
for almost all $t\in \R$. If we pick a sequence $q_k \to \infty$ 
we can choose the exceptional set of measure zero where (\ref{qchargeconsae}) 
does not hold uniformly in $k$, and passing to the limit shows that the 
$L^\infty$-norm of $f(t)$ is preserved in the same sense.

Combining this with the weak continuity of the map $t\mapsto f(t)$
or rather a suitable representative, the Radon-Riesz Theorem implies
that for every $q\in ]1,\infty[$ one can choose a set of continuity
$C_q$ with the asserted properties, and via two sequences 
$q_k^+ \to \infty$ and $q_k^- \searrow 1$ we obtain a set of continuity
$C$ which works for all $q\in ]1,\infty[$ simultaneously. 
\hspace*{\fill} $\Box$

Note that the relativistic nature of the system was exploited
in the proof above. We are not aware of a proof of 
(\ref{chargeconsae}) in the non-relativistic case. 

\section{The result---statement and comments} 
\setcounter{equation}{0}

Collecting the results of the previous sections we arrive
at the following existence result for global weak solutions
to the relativistic Vlasov-Maxwell system:

\smallskip

\noindent
{\bf Theorem.}
{\em For initial data $\open{f}\, \in L^1_\mathrm{kin}\cap L^\infty (\R^6),\ 
\open{E}\,,\ \open{B}\, \in L^2(\R^3)$
which satisfy the constraints (\ref{constr}) in the sense of distributions
there exists a global weak solution of the relativistic Vlasov-Maxwell system,
i.e., there exist functions 
\[
f\in L^\infty(\R;L^1_\mathrm{kin}\cap L^\infty (\R^6)),
\ E, B \in L^\infty (\R; L^2(\R^3)),\
\rho, j \in  L^\infty (\R; L^{4/3}(\R^3))
\]
such that $(f,E,B)$ satisfy (\ref{vlasov})---(\ref{constr}) in the
sense of distributions with $\rho,\;j$ defined in terms of $f$ by
(\ref{rhojdef}). The function $f$ is a.~e.\ non-negative.

The mappings $t \mapsto f(t),E(t),B(t)$ are (after choosing suitable
representatives) continuous with respect to the following topologies:
the standard topology in the space of distributions ${\cal D}'(\R^6)$
or ${\cal D}'(\R^3)$ respectively,
the weak topology of $L^2$, and the strong topology of $W^{-s,2}(M)$
for any $s>0$ and any bounded measurable subset $M$ of $\R^6$
or $\R^3$ respectively. The mapping $t\mapsto f(t)$ is a.~e.\ strongly
continuous into any $L^q(\R^6)$, $1<q<\infty$, 
in the sense of Proposition~\ref{lpcons}.
The initial conditions (\ref{incond}) are satisfied.

At a.~e.~time $t$ the total energy
\[
\int\!\!\!\!\!\int \sqrt{1+|p|^2} f(t,x,p)\,dp\,dx + 
\frac{1}{8 \pi} \int \left(|E(t,x)|^2 + |B(t,x)|^2\right)\, dx
\]
is bounded by its value at $t=0$. The total charge is conserved,
\[
\int\!\!\!\!\!\int f (t)\, dp\,dx =  \int\!\!\!\!\!\int \open{f}\, \, dp\,dx\
\mbox{for a.~a.}\ t\in \R,
\] 
and the same is true for any $L^q$-norm of $f(t)$, $1\leq q \leq \infty$.
}

\smallskip

\noindent
The following deficiencies of weak solutions are obvious: Firstly,
uniqueness is not known. Secondly, it is not clear whether
energy is conserved and whether $L^q$-norms of $f$ are conserved
{\em everywhere}.
Moreover, their continuity with respect to $t$ holds either
everywhere in a rather weak sense or only a.~e.\ in the strong $L^q$ sense.
A less obvious disadvantage is the following:
It is not clear that any
weak solution in the sense of the theorem can be obtained
via the particular regularization which we employed here,
and it is conceivable that results for weak solutions
depend on the way in which these are constructed.

\section*{Appendix: Momentum averaging}
\setcounter{equation}{0}
To prove Lemma~\ref{momav} we let $\psi \in C^\infty_c(\R^3)$ 
with $\mathrm{supp}\, \psi \subset B_R$
and $R>0$. By the usual definition of the fractional order Sobolev
spaces via Fourier transforms,
\begin{equation} \label{h1/4est}
\left\|\int h(\cdot,\cdot,p)\,\psi(p)\, dp\right\|^2_{H^{1/4}}
\leq
\|\psi\|_2^2 \|\hat h\|_2^2 + \int\!\!\!\!\!\int
\left| I(\tau,\xi) \right|^2 
(|\tau|^{1/2} + |\xi|^{1/2})\, d\xi\, d\tau
\end{equation}
where 
\begin{equation} \label{idef}
I(\tau,\xi) := \int \hat h (\tau,\xi,p)\,\psi(p)\, dp
\end{equation}
and $\hat h$ denotes the Fourier transform of $h$ with respect to
$(t,x)$. Confusion with the notation $\widehat{p}$ seems unlikely. 
By assumption, $h$ satisfies the transport equation (\ref{transport})
on $\R \times \R^6$ in the sense of distributions, and hence
\begin{equation} \label{vlft}
i (\tau + \widehat{p} \cdot \xi) \hat h = \hat g_0  + \mathrm{div}_p \hat g_1.
\end{equation}
This identity is only useful where
the factor on the left hand side is away from zero. Hence we let
$\zeta \in C^\infty_c(\R)$ be such that
\[
0\leq \zeta \leq 1,\ \mathrm{supp}\, \zeta \subset[-2,2],\ \zeta_{|[-1,1]}=1 
\]
and we split the integral (\ref{idef}) into the two parts
\begin{eqnarray*}
I_1 (\tau,\xi) 
&:=& 
\int \hat h (\tau,\xi,p)\,\psi(p)\, 
\zeta\left(\frac{\tau + \widehat{p} \cdot \xi}{\kappa}\right)\,dp,\\
I_2 (\tau,\xi) 
&:=& 
\int \hat h (\tau,\xi,p)\,\psi(p)\, 
\left[1-\zeta\left(\frac{\tau + \widehat{p} \cdot \xi}{\kappa}
\right)\right]\,dp,
\end{eqnarray*}
where $\kappa >0$ will be chosen appropriately in dependence of $\xi$.

For almost all $(\tau,\xi) \in \R^4$,
\begin{equation} \label{I1est}
|I_1 (\tau,\xi)| \leq C \|\hat h(\tau,\xi,\cdot)\|_2 
\left(\frac{\kappa}{|\xi|}\right)^{1/2} 
{\bf 1}_{|\tau|\leq r |\xi| + 2 \kappa},
\end{equation}
where $r := R/\sqrt{1+R^2}$---as in the rest of the argument,
the $L^2$-norm with respect to $p$ refers to the ball $B_R$ with $R>0$ fixed,
and ${\bf 1}$ with some subscript is the ``indicator function''
of the set or condition in the subscript. 
To see (\ref{I1est}) we apply the Cauchy-Schwarz
inequality and observe that
\[
\int_{B_R}\zeta^2 \left(\frac{\tau + \widehat{p} \cdot \xi}{\kappa}\right)\,dp
\leq
\int_{B_R}{\bf 1}_{[-\tau -2\kappa,-\tau + 2\kappa]}(\widehat{p}_1 |\xi|)\, dp;
\]
without loss of generality we may assume that $\xi = (|\xi|,0,0)$.
The integral on the right hand side vanishes if 
$|\tau| > r |\xi| + 2 \kappa$, and via a change of variables it can 
easily be estimated against $\kappa/|\xi|$ which proves (\ref{I1est}).

For the estimate of $I_2$ the basic idea is to use
(\ref{vlft}) and to integrate by parts in the term containing 
the divergence of $g_1$. With the abbreviation
\[
\chi(\tau,\xi,p) := \psi(p) 
\left[1-\zeta\left(\frac{\tau + \widehat{p} \cdot \xi}{\kappa}\right)\right]
\frac{1}{\tau + \widehat{p} \cdot \xi}
\]
it follows that
\begin{equation} \label{I2rep}
I_2(\tau,\xi) = - i
\left(\int \hat g_0(\tau,\xi,p)\,\chi(\tau,\xi,p)\,dp
- \int \hat g_1(\tau,\xi,p)\cdot\partial_p\chi(\tau,\xi,p)\,dp\right)
\end{equation}
where this last identity holds a.~e.\ on $\R \times \R^3$ if $\kappa = 1$
and a.~e.\ on $\R \times \{|\xi|>1\}$ if $\kappa = |\xi|^{1/2}$.
The first choice will be used for small $|\xi|$ and the second one
for large $|\xi|$. For any multi-index $\alpha \in \N^3$ with length
$|\alpha|\leq 1$,
\[
|D_p^\alpha \chi(\tau,\xi,p)| \leq C {\bf 1}_{|p|\leq R} 
{\bf 1}_{|\tau + \widehat{p}\cdot \xi| \geq \kappa} 
\frac{1}{|\tau + \widehat{p} \cdot \xi|}
\left(1+\frac{|\xi|}{\kappa}\right).
\]
Hence
\[
\|D^\alpha_p \chi(\tau,\xi,\cdot)\|_2 \leq C 
\left(1+\frac{|\xi|}{\kappa}\right) \frac{1}{|\xi|^{1/2}}
\left(\int_{\tau-r |\xi|}^{\tau+r |\xi|} {\bf 1}_{|w|\geq \kappa}
\frac{1}{w^2}dw\right)^{1/2} .
\]
The remaining integral is straightforwardly estimated by $C/\kappa$.
In addition, for $|\tau|\geq r |\xi| + 2\kappa$ the origin 
does not lie in the
domain of integration of this integral, and we can estimate it by
integrating $1/w^2$ over the domain. Hence,
\[
\|D^\alpha_p \chi(\tau,\xi,\cdot)\|_2
\leq
C \left(1+\frac{|\xi|}{\kappa}\right)
\left({\bf 1}_{|\tau|\leq r |\xi| + 2 \kappa}\frac{1}{\kappa|\xi|}
+ {\bf 1}_{|\tau| > r |\xi| + 2 \kappa} \frac{1}{\tau^2 -r^2|\xi|^2}
\right)^{1/2} .
\]
Let us abbreviate
\[
N(\tau,\xi):= \|\hat h (\tau,\xi,\cdot)\|_2 + \|\hat g_0 (\tau,\xi,\cdot)\|_2
+ \|\hat g_1 (\tau,\xi,\cdot)\|_2.
\]
Combining (\ref{I1est}) with (\ref{I2rep}) and the estimate for 
$D^\alpha_p \chi$ we finally obtain the following estimate for $I$,
\begin{eqnarray*}
|I(\tau,\xi)|
&\leq&
C\, N(\tau,\xi)\, \biggl[ {\bf 1}_{|\tau|\leq r |\xi| + 2 \kappa}
\left(\frac{\kappa}{|\xi|} + \left(1+\frac{|\xi|}{\kappa}\right)^2
\frac{1}{\kappa|\xi|}\right) \\
&&
\qquad \qquad \quad
+ {\bf 1}_{|\tau|\geq r |\xi| + 2 \kappa}
\left(1+\frac{|\xi|}{\kappa}\right)^2
\frac{1}{\tau^2 -r^2|\xi|^2}\biggr]^{1/2},
\end{eqnarray*}
which holds a.~e.\ on $\R \times \R^3$ if $\kappa = 1$
and a.~e.\ on $\R \times \{|\xi|>1\}$ if $\kappa = |\xi|^{1/2}$.
Hence for $|\xi|\leq 1$ we take $\kappa = 1$ to obtain
\[
|I(\tau,\xi)|
\leq
C\, N(\tau,\xi)\, \left( {\bf 1}_{|\tau|\leq r |\xi| + 2}
\frac{1}{|\xi|} + 
{\bf 1}_{|\tau|\geq r |\xi| + 2}
\frac{1}{\tau^2 -r^2|\xi|^2}\right)^{1/2},
\]
and for $|\xi| > 1$ we take $\kappa = |\xi|^{1/2}$ to obtain
\[
|I(\tau,\xi)|
\leq
C\, N(\tau,\xi)
\left({\bf 1}_{|\tau|\leq r |\xi| + 2|\xi|^{1/2}}
\frac{1}{|\xi|^{1/2}} + 
{\bf 1}_{|\tau|\geq r |\xi| + 2|\xi|^{1/2}}
\frac{|\xi|}{\tau^2 -r^2|\xi|^2}\right)^{1/2}.
\]
Now we split the right hand side of (\ref{h1/4est}) as follows:
\begin{eqnarray*}
&&
\int\!\!\!\!\!\int|I(\tau,\xi)|^2(|\tau|^{1/2} + |\xi|^{1/2})\,d\tau\,d\xi 
= \int\!\!\!\!\!\int|I|^2 |\tau|^{1/2} {\bf 1}_{|\xi|>1} d\tau\,d\xi\\
&&\qquad
+ \int\!\!\!\!\!\int|I|^2 |\tau|^{1/2} {\bf 1}_{|\xi|\leq 1} 
{\bf 1}_{|\tau|>r + 2} d\tau\,d\xi
+ \int\!\!\!\!\!\int|I|^2 |\tau|^{1/2} {\bf 1}_{|\xi|\leq 1} 
{\bf 1}_{|\tau|\leq r + 2} d\tau\,d\xi\\
&&\qquad
+ \int\!\!\!\!\!\int|I|^2 |\xi|^{1/2} {\bf 1}_{|\xi|\leq 1} d\tau\,d\xi
+ \int\!\!\!\!\!\int|I|^2 |\xi|^{1/2} {\bf 1}_{|\xi| > 1} d\tau\,d\xi\\
&&
=: A_1 + A_2 + A_3 + A_4 + A_5.
\end{eqnarray*}
Using the appropriate part of the estimate for $I$ we find that
\[
A_1 + A_2 + A_5 \leq C \int\!\!\!\!\!\int N^2\,d\tau\,d\xi
= C(\|\hat h \|_2^2 + \|\hat g_0 \|_2^2
+ \|\hat g_1\|_2^2)
\]
as desired, while the terms $A_3$ and $A_4$ can be estimated directly
by $\|\hat h (\tau,\xi,\cdot)\|_2^2$ via the Cauchy-Schwarz inequality,
and the proof of Lemma~\ref{momav} is complete.

\smallskip

{\bf Acknowledgment:}
The present notes originate in a seminar on kinetic theory
at the University of Vienna. 
I would like to thank the participants 
of that seminar as well as Prof.~R.~T.~Glassey for their interest 
in the simplifications I report here.


\begin{thebibliography}{10}


\bibitem{BG}
F.~Bouchut, F.~Golse, C.~Pallard:
On classical solutions to the 3d relativistic Vlasov-Maxwell
system: Glassey-Strauss' theorem revisited.
Preprint, 2003, arXiv:math.AP/0301175v1 

\bibitem{CR}
S.~Calogero, G.~Rein:
Global weak solutions to the Nordstr\"om-Vlasov system.
{\em J.~Differential Eqns.}, to appear

\bibitem{DL1} 
R.~J.~DiPerna, P.-L.~Lions: 
Global weak solutions of Vlasov-Maxwell systems. 
{\em Commun.\ Pure Appl.\ Math.}\ 
{\bf 42}, 6,
729--757 (1989)
 
\bibitem{DL2} 
R.~J.~DiPerna, P.-L.~Lions: 
On the Cauchy problem for Boltzmann equations; global existence and 
weak stability.
{\em Ann.\ Math.}\ 
{\bf 130}, 321--366 (1989) 

\bibitem{Gl1}
R.~T.~Glassey: {The Cauchy Problem in Kinetic Theory},
SIAM, Philadelphia (1996)

\bibitem{GlSch}
R.~T.~Glassey, J.~Schaeffer:
On symmetric solutions of the relativistic Vlasov-Poisson system.
{\em Commun.\ Math.\ Phys.}\
{\bf 101}, 459--473 (1985)

\bibitem{GlSch1}
R.~T.~Glassey, J.~Schaeffer:
Global existence for the relativistic Vlasov-Maxwell system with nearly
neutral data.
{\em Commun.\ Math.\ Phys.}\
{\bf 119}, 353--384 (1988)

\bibitem{GlSch2}
R.~T.~Glassey, J.~Schaeffer:
Control of velocities generated in a two-dimensional collisionless plasma
with symmetry.
{\em Transp.\ Theory and Stat.\ Mech.}\
{\bf 17}, 467--560 (1988)

\bibitem{GlSch3}
R.~T.~Glassey, J.~Schaeffer:
On the ``one and one-half dimensional'' relativistic Vlasov-Maxwell
system.
{\em Math.\ Meth.\ Appl.\ Sci.}\
{\bf 13}, 169--179 (1990)

\bibitem{GlSch4}
R.~T.~Glassey, J.~Schaeffer:
The ``two and one-half dimensional'' relativistic Vlasov-Maxwell
system.
{\em Commun.\ Math.\ Phys.}\
{\bf 185}, 257--284 (1997)

\bibitem{GlSch5}
R.~T.~Glassey, J.~Schaeffer:
The relativistic Vlasov-Maxwell system
in two space dimensions: Parts I \& II.
{\em Arch.\ Rational Mech.\ Anal.}\
{\bf 141}, 331--354 \& 355--374 (1998)

\bibitem{GlSt1}
R.~T.~Glassey, W.~Strauss:
Singularity formation in a
collisionless plasma could occur only at high velocities. 
{\em Arch.\ Rat.\ Mech.\ Anal.}\ 
{\bf 92}, 59--90 (1986)

\bibitem{GlSt2}
R.~T.~Glassey, W.~Strauss:
Absence of shocks in an initially dilute collisionless plasma.
{\em Commun.\ Math.\ Phys.}\ 
{\bf 113}, 191--208 (1987)

\bibitem{GLPS}
F.~Golse, P.-L.~Lions, B.~Perthame, R.~Sentis:  
Regularity of the moments of the solution of a transport equation. 
{\em J. Funct. Anal.} {\bf 76}, no.\ 1, 110--125 (1988).

\bibitem{Ho}
E.~Horst:
Global solutions of the relativistic Vlasov-Maxwell system of plasma physics.
{\em Dissertationes Mathematicae}
{\bf CCXCII}, 1--63 (1990)

\bibitem{KS}
S.~Klainerman, G.~Staffilani:
A new approach to study the Vlasov-Maxwell system.
{\em Commun.\ Pure Appl.\ Anal.}\
{\bf 1}, 103--125 (2002)

\bibitem{Kr}
K.~Kruse:
{\em Ein neuer Zugang zur globalen Existenz von Distributionen\-l\"o\-sungen
des Vlasov-Maxwell-Systems partieller Differentialgleichungen.}
Diploma thesis, University of Munich, 1991

\bibitem{KR} 
K.~Kruse, G.~Rein: 
A stability result for the relativistic Vlasov-Maxwell system. 
{\em Arch.\ Rational Mech.\ Anal.}\ 
{\bf 121}, 2, 187--203 (1992)

\bibitem{KRST} 
M.~Kunzinger, G.~Rein, R.~Steinbauer, G.~Teschl: 
Global weak solutions of the relativistic Vlasov-Klein-Gordon
system. 
{\em Commun.\ Math.\ Phys.}\ {\bf 238}, 1-2, 367--378 (2003)

\bibitem{LP}
P.-L.~Lions, B.~Perthame:
Propagation of moments and regularity for the 3-dimensional 
Vlasov-Poisson system.
{\em Invent.\ Math}.\ 
{\bf 105}, 415--430 (1991)

\bibitem{Pf}
K.~Pfaffelmoser: 
Global classical solutions of the Vlasov-Poisson system in three dimensions 
for general initial data.
{\em J.\ Diff.\ Eqns}.\ 
{\bf 95} (1992), 281--303.

\bibitem{R1}
G.~Rein: 
Generic global solutions of the relativistic Vlasov-Maxwell
system of plasma physics. 
{\em Commun.\ Math.\ Phys.}\ 
{\bf 135}, 41--78 (1990)

\bibitem{Sch} 
J.~Schaeffer: 
Global existence of smooth solutions to the Vlasov-Poisson system 
in three dimensions.
{\em Commun.\ Part.\ Diff.\ Eqns.}\ 
{\bf 16} (1991), 1313--1335.


\end{thebibliography}
\end{document}